\documentclass[11pt]{article}
\usepackage{graphicx}
\usepackage{longtable,lscape}
\usepackage{amsfonts}
\usepackage{float}
\newcommand{\compl}{{\mathbb C}}

\newcommand{\captionfonts}{\footnotesize}
\makeatletter  
\long\def\@makecaption#1#2{%
  \vskip\abovecaptionskip
  \sbox\@tempboxa{{\captionfonts #1: #2}}%
  \ifdim \wd\@tempboxa >\hsize
    {\captionfonts #1: #2\par}
  \else
    \hbox to\hsize{\hfil\box\@tempboxa\hfil}%
  \fi
  \vskip\belowcaptionskip}
\makeatother   
\begin{document}
\title{Interpreting Quantum Particles as Conceptual Entities}
\author{Diederik Aerts\\
        \normalsize\itshape
        Center Leo Apostel for Interdisciplinary Studies \\
        \normalsize\itshape
        and Departments of Mathematics and Psychology \\
        \normalsize\itshape
        Vrije Universiteit Brussel, 1160 Brussels, 
       Belgium \\
        \normalsize
        Email: \textsf{diraerts@vub.ac.be}
        }
\date{}
\maketitle              
\begin{abstract}
\noindent We elaborate an interpretation of quantum physics founded on the hypothesis that quantum particles are conceptual entities playing the role of communication vehicles between material entities composed of ordinary matter which function as memory structures for these quantum particles. We show in which way this new interpretation gives rise to a natural explanation for the quantum effects of interference and entanglement by analyzing how interference and entanglement emerge for the case of human concepts. We put forward a scheme to derive a metric based on similarity as a predecessor for the structure of `space, time, momentum, energy' and `quantum particles interacting with ordinary matter' underlying standard quantum physics, within the new interpretation, and making use of aspects of traditional quantum axiomatics. More specifically, we analyze how the effect of non-locality arises as a consequence of the confrontation of such an emerging metric type of structure and the remaining presence of the basic conceptual structure on the fundamental level, with the potential of being revealed in specific situations. 
\end{abstract}
\section{Introduction}
Inspired by our earlier use of the quantum mechanical formalism to model conceptual situations as they appear in cognition, decision theory and economics \cite{aerts2009a,aerts2009b,aerts2009c,aertsaertsgabora2009,aertsdhooghe2009,aerts2007a,aerts2007b,aertsczachordhooghe2006,aertsgabora2005a,aertsgabora2005b,aertsczachor2004,gaboraaerts2002,aertsbroekaertsmets1999a,aertsbroekaertsmets1999b,aertsaerts1994}, and drawing on former insights on quantum physics \cite{aertsaertsbroekaertgabora2000,aerts1999a,aerts1999b,aerts1998a,aerts1998b,aerts1995,aerts1994,aerts1993,aerts1992,aerts1986}, we put forward a new view that we intend to elaborate into a full interpretation of quantum theory. Our view proposes an answer to the question of `What is a quantum entity?' in a way that is different from what is suggested by existing interpretations, giving rise to an alternative approach to specific and well-known issues in quantum theory. This article focuses on fundamental aspects of quantum theory, such as interference, entanglement, quantum identity and the measurement problem, and we refer to \cite{aerts2009a} for a more detailed exposition of this view in a more global perspective. The new view that we put forward also generates a radically different way of considering the `reality' described by quantum physics, i.e. the reality of the micro-world, as well as its relation to the macro-world. And it has fundamental consequences for our understanding of aspects of the ordinary everyday world. 

\section{The nature of a quantum entity}
From the birth of quantum theory, one of the fundamental questions has been: `What does the quantum wave function represent?' There are several reasons why, even today, an answer to this question cannot readily be given. One of these, commonly regarded to be the most important, is often referred to as `the measurement problem'. The effect of a measurement on a quantum entity is described by a change of the wave function, often called `collapse', so that the question can be rephrased as: `What is the meaning of this collapse?'

In the early years, before some of the aspects of quantum mechanics were properly understood, the collapse was often considered to be an effect on our knowledge about the quantum entity. The measurement was thought to make this knowledge more specific and hence to affect its representation. One consequence was that the wave function itself would represent our knowledge of the quantum entity. It also meant that the wave function would not describe the reality of the quantum entity, so that a collapse would not affect this reality but only our knowledge of this reality. Let us call this the `knowledge view'. For a number of reasons, the `knowledge view' has never been regarded as an adequate interpretation of quantum theory. One important reason is the existence of an effect considered to be the most crucial of all quantum effects, namely `interference'. To clarify this, we will consider the archetypical double-slit interference situation with two slits $A$ and $B$ \cite{tonomuraendomatsudakawasakiezawa1989,jonsson1961,young1802}. The interference that occurs in the double-slit situation is strictly linked to the value of the individual wave functions $\psi_A(x,y,z)$ and $\psi_B(x,y,z)$ behind each of the slits, because it is `the normalized sum of these individual wave functions', i.e. expression (\ref{sumwavefunction}), that constitutes the `total wave function' behind the slits. The integral over a specific region of space $R$ of the square of the absolute value of the normalized sum wave function, which is expressions (\ref{withinterference}) and (\ref{withinterferenceworkedout}), defining the probability of detection
\begin{eqnarray} \label{sumwavefunction}
&&{1 \over \sqrt{2}}(\psi_A(x,y,z)+\psi_B(x,y,z)) \\
\label{withinterference}
&&{1 \over 2}\int_R|\psi_A(x,y,z)+\psi_B(x,y,z)|^2dxdydz \\ \label{withinterferenceworkedout}
&&={1 \over 2}\int_R(|\psi_A(x,y,z)|^2+|\psi_B(x,y,z)|^2+2\Re(\psi_A(x,y,z)^*\psi_B(x,y,z)))dxdydz  \\
\label{withoutinterference}
&&{1 \over 2}\int_R(|\psi_A(x,y,z)|^2+|\psi_B(x,y,z)|^2)dxdydz \\ \label{interferenceterm}
&&\int_R\Re(\psi_A(x,y,z)^*\psi_B(x,y,z))dxdydz
\end{eqnarray}
in this region $R$, is different from the average of the integral over this region $R$ of the squares of the absolute values of the individual wave functions, which is expression (\ref{withoutinterference}),
defining the average of the probabilities of detection for each of the individual slits in region $R$. Interference is the effect manifesting this difference, so that it is modeled by the expression (\ref{interferenceterm}), which is (\ref{withinterferenceworkedout}) minus (\ref{withoutinterference}).
One of many physicists who have analyzed this situation in depth is Richard Feynman, who gave a detailed account of the resulting weird effects \cite{feynman1970,feynman1965}. Apart from the possibility of explaining these effects, it is clear that it is the wave function rather than the square of its absolute value that describes what happens. It also shows that the knowledge view does not suffice as an adequate interpretation of quantum theory.

As a consequence, the wave function came to be interpreted as representing the reality of the quantum entity itself. Although such an interpretation properly explains the effect of interference -- `quantum waves interfere in a similar way as real waves interfere in the macro-world' --, it immediately raises a multitude of new problems. For example, the question of `What is the effect that corresponds to the collapse' can then be answered only by saying that `the collapse is a real effect of change of the reality of the quantum entity itself', which means that the collapse changes the pure state of a quantum entity into another pure state. However, if the collapse is an effect of change of the reality of the quantum entity, it should in principle be possible to put forward an evolution equation offering a dynamic description of the collapse. There are substantial problems connected to constructing such an evolution equation \cite{pearle2007,gisin1989,ghirardiriminiweber1986}, and besides, experiments suggest that during the collapse something seems to be happening at a speed far beyond the speed of light \cite{zbindenbrendeltittelgisin2001}. Furthermore, the `no signal faster than light' hypothesis of relativity theory limits considerably the possible types of collapse equations \cite{simonbuzekgisin2001,svetlichny1998,gisin1989}. These substantial difficulties with dynamical collapse theories have contributed to a growing interest in what was originally considered a rather implausible type of quantum interpretation, namely the many-world interpretation \cite{deutsch1999,dewittgraham1973,everett1957}. The many-world interpretation is a radical interpretation in that its main hypothesis involves a negation of the collapse as a real physical process while supporting the notion that the wave function represents the reality of the quantum entity. The main idea is that there is no collapse and that all possible outcomes actually evolve in parallel universes, and hence exist as realities. The many-world interpretation thus regards `the possible' as `the real', and asserts that we human beings are made to believe that `the possible' is not `the real', because we are locked inside one of the parallel universes. The many-world interpretation is spectacular and has won some strong support. In the early days, we shared the widespread skepticism about this view and today we still maintain a very critical opinion about it. We will discuss this in further detail in future work where we will confront the many-world interpretation with the new interpretation developed in the present article.

A well-known and extensively tested and confirmed effect of quantum theory that poses even more problems when interpreted according to the view that `the wave function represents the reality of the quantum entity', is that of the entanglement of different quantum entities \cite{aspectgrangierrobert1981,bell1964,einsteinpodolskyrosen1935}. If we consider two quantum entities, the wave function of the compound entity consisting of these two quantum entities is a complex function of six real variables, which, in case there is entanglement, is `not' the product of two functions of three real variables. This is what originates the mysterious correlations that give rise to the effect of quantum non-locality. The spectacular nature of quantum non-locality is very much apparent from a recent experiment with two entangled photons traveling to different locations 18 km apart which measured the correlations produced by their entanglement \cite{salartbaasbranciardgisinzbinden2008}. We will see that entanglement and non-locality are both incorporated, explained and understood in a natural way in our new interpretation.

Although we now realize that earlier research had sown the seeds for many aspects of this new interpretation \cite{aertsaertsbroekaertgabora2000,aerts1999a,aerts1998a,aerts1998b,aerts1995,aerts1994,aerts1992,aerts1986}, it did not reach full bloom until our recent findings on the modeling of concepts and combinations of concepts by means of the quantum mechanical formalism \cite{aerts2009a,aerts2009b,aerts2009c,aertsaertsgabora2009,aertsdhooghe2009,aerts2007a,aerts2007b,aertsczachordhooghe2006,aertsgabora2005a,aertsgabora2005b,aertsczachor2004,gaboraaerts2002,aertsaerts1994}. The main aspect of the new interpretation we propose in the present article is a specific hypothesis about the nature of a quantum entity itself.

\bigskip
\noindent
Hypothesis {\it NQE} (nature quantum entity): {\it The nature of a quantum entity is `conceptual', i.e. it interacts with a measuring apparatus (or with an entity made of ordinary matter) in an analogous way as a concept interacts with a human mind (or with an arbitrary memory structure sensitive to concepts).}

\bigskip
\noindent
Hypothesis {\it NQE} means that the quantum entity affects a measuring apparatus (or an entity made of ordinary matter) like a concept affects a human mind (or an arbitrary memory structure sensitive to concepts). It therefore also means that the quantum entity affects a measuring apparatus or an entity made of ordinary matter `not' like an object affects another object. We could have formulated hypothesis {\it NQE} by stating that a quantum entity `is a concept' and `not an object'. However, we prefer to use the notion of `conceptual' as the notion that can give rise to `the being' of a quantum entity as well as to `the being' of a human concept, in the same way that the notion of `wave' gives rise to `the being' of an electromagnetic wave but also to `the being' of a sound wave.

In a sense, this new interpretation is in between what we have called the knowledge view and the common interpretations that consider the quantum entity to be an object. Indeed, on the one hand a concept behaves like a piece of knowledge with respect to the memory structure that is sensitive to it, but on the other hand it also has object-like aspects and it can be treated as an object in intersubjective terms, where `intersubjective' is defined with respect to the memory structures involved. By introducing our basic hypothesis {\it NQE}, we assign a special role to the measuring apparatus with respect to the quantum entity, but we do this in a `relative sense', i.e. measuring apparatuses -- and more generally entities made of ordinary matter -- are to quantum entities what memory structures -- be they human or artificial -- are to human concepts. We will analyze in the following what the hypothesis {\it NQE} implies for other aspects of the world, for example for the nature of macroscopic material entities composed of ordinary matter. However, we first need to explain what we mean by interference between human concepts.

\section{Quantum interference}
In this section we will present a number of aspects of our research on the quantum modeling of human concepts \cite{aerts2009a,aerts2009b,aerts2009c,aertsaertsgabora2009,aertsdhooghe2009,aerts2007a,aerts2007b,aertsczachordhooghe2006,aertsgabora2005a,aertsgabora2005b,aertsczachor2004,gaboraaerts2002,aertsaerts1994} that are relevant to our new interpretation of quantum mechanics. We will consider the same two concepts throughout our discussion, viz. {\it Fruits} and {\it Vegetables}, to illustrate the theoretical aspects of our quantum modeling of concepts.

To begin with, we use the term `states' to describe what psychologists call `exemplars' or `instantiations' of a concept 
\cite{aertsgabora2005a}. In our example, {\it Apple, Raisin, Coconut} and {\it Elderberry}, which in psychology are seen as instances or exemplars of the concept {\it Fruits}, are considered states of the entity corresponding to the concept {\it Fruits}, while {\it Broccoli, Pumpkin, Tomato} and {\it Green Pepper} are considered states of the entity corresponding to the concept {\it Vegetables}. In \cite{aertsgabora2005b} we showed that the entity corresponding to a concept can be modeled by the quantum mechanical formalism, i.e. with its `states' represented by unit vectors in a complex Hilbert space, and the `probabilities of change of state' modeled by quantum probabilities. Our earlier thoughts on the non-classical quantum-like nature of human decision processes  \cite{aertsaerts1994} and the structure of concept combinations \cite{gaboraaerts2002} inspired us to elaborate such a quantum model for concepts. The strong similarities between successful and well-known cognitive science formalisms such as `Latent Semantic Analysis', other vector space based semantic formalisms \cite{bleingmichaeljordan2003,griffithssteyvers2002,hofmann1999,lundburgess1995,deerwesterdumaisfurnaslandauerharshman1990}, and the quantum formalism, convinced us that there was a profound structural connection \cite{aertsczachor2004}, which laid the foundation for the quantum model we developed for concepts and their combinations \cite{aertsgabora2005a,aertsgabora2005b}. At first, we believed that the capacity of quantum structures to model concepts was due to the highly contextual nature of concept combinations and the natural way in which quantum mechanics models contextuality \cite{gaboraaerts2002,aertsgabora2005a,aertsgabora2005b}. Later, however, we observed that also the effect of interference as we know it in quantum mechanics occurs naturally when concepts are combined. Hence, although concepts are little packets of knowledge with respect to the memory structure that is sensitive to them, they entail interference effects when combined, because, as we will see, they are concepts by nature and constantly originate new concepts \cite{aerts2009a,aerts2009b,aerts2009c,aerts2007a,aerts2007b}. This is what made us believe that our findings would prove valuable to the understanding of quantum mechanics itself and could inspire a new interpretation.

For a better understanding of the occurrence of interference when concepts are combined, we will now briefly analyze one of the specific situations studied at greater length in \cite{aerts2009a,aerts2009b}. Consider the concepts {\it Fruits} and {\it Vegetables} and a list of exemplars described as states of both concepts, more specifically the 24 exemplars in Table 1. We consider the following experimental situation: Human beings -- `subjects', in the terminology of psychology -- are asked the following three questions: {\it Question $A$}: `Choose one of the exemplars of the list of Table 1 that you find a good example of {\it Fruits}'. {\it Question $B$}: `Choose one of the exemplars of the list of Table 1 that you find a good example of {\it Vegetables}'. {\it Question $A$ or $B$}: `Choose one of the exemplars of the list of Table 1 that you find a good example of {\it Fruits or Vegetables}'. Then we calculate the relative frequencies $\mu(A)_k$, $\mu(B)_k$ and $\mu(A\ {\rm or}\ B)_k$, i.e. the number of times that exemplar $k$ is chosen divided by the total number of choices made to the three questions $A$, $B$ and $A\ {\rm or}\ B$, respectively, and interpret the outcome as an estimate for the probabilities that exemplar $k$ is chosen for respective questions $A$, $B$ and $A\ {\rm or}\ B$.
These relative frequencies are given in Table 1. For example, for {\it Question $A$}, of 10,000 subjects, 359 chose {\it Almond}, hence $\mu(A)_1=0.0359$, 425 chose {\it Acorn}, hence $\mu(A)_2=0.0425$, 372 chose {\it Peanut}, hence $\mu(A)_3=0.0372$, $\ldots$, and 127 chose {\it Black Pepper}, hence $\mu(A)_{24}=0.0127$. Analogously for {\it Question $B$}, of 10,000 subjects, 133 chose {\it Almond}, hence $\mu(B)_1=0.0133$, 108 chose {\it Acorn}, hence $\mu(B)_2=0.0108$, 220 chose {\it Peanut}, hence $\mu(B)_3=0.0220$, $\ldots$, and 294 chose {\it Black Pepper}, hence $\mu(B)_{24}=0.0294$, and for {\it Question $A\ {\rm or}\ B$}, 269 chose {\it Almond}, hence $\mu(A\ {\rm or}\ B)_1=0.0269$, 249 chose {\it Acorn}, hence $\mu(A\ {\rm or}\ B)_2=0.249$, 269 chose {\it Peanut}, hence $\mu(A\ {\rm or}\ B)_3=0.269$, $\ldots$, and 222 chose {\it Black Pepper}, hence $\mu(A\ {\rm or}\ B)_{24}=0.222$.
\begin{table}[htb]
\footnotesize
\begin{center}
\begin{tabular}{|llllllll|}
\hline 
\multicolumn{2}{|l}{} & \multicolumn{1}{l}{$\mu(A)_k$} & \multicolumn{1}{l}{$\mu(B)_k$} & \multicolumn{1}{l}{$\mu(A\ {\rm or}\ B)_k$} & \multicolumn{1}{l}{${\mu(A)_k+\mu(B)_k \over 2}$} & \multicolumn{1}{l}{$\lambda_k$} & \multicolumn{1}{l|}{$\theta$} \\
\hline
\multicolumn{8}{|l|}{\it $A$=Fruits, $B$=Vegetables} \\
\hline
1 & {\it Almond} & 0.0359 & 0.0133 & 0.0269 & 0.0246 & 0.0218 & 83.8854$^\circ$ \\
2 & {\it Acorn} & 0.0425 & 0.0108 & 0.0249 & 0.0266 & -0.0214 & -94.5520$^\circ$ \\
3 & {\it Peanut} & 0.0372 & 0.0220 & 0.0269 & 0.0296 & -0.0285 & -95.3620$^\circ$ \\
4 & {\it Olive} & 0.0586 & 0.0269 & 0.0415 & 0.0428 & 0.0397 & 91.8715$^\circ$ \\
5 & {\it Coconut} & 0.0755 & 0.0125 & 0.0604 & 0.0440 & 0.0261 & 57.9533$^\circ$ \\
6 & {\it Raisin} & 0.1026 & 0.0170 & 0.0555 & 0.0598 & 0.0415 & 95.8648$^\circ$ \\
7 & {\it Elderberry} & 0.1138 & 0.0170 & 0.0480 & 0.0654 & -0.0404 & -113.2431$^\circ$ \\ 
8 & {\it Apple} & 0.1184 & 0.0155 & 0.0688 & 0.0670 & 0.0428 & 87.6039$^\circ$ \\ 
9 & {\it Mustard} & 0.0149 & 0.0250 & 0.0146 & 0.0199 & -0.0186 & -105.9806$^\circ$ \\
10 & {\it Wheat} & 0.0136 & 0.0255 & 0.0165 & 0.0195 & 0.0183 & 99.3810$^\circ$ \\ 
11 & {\it Root Ginger} & 0.0157 & 0.0323 & 0.0385 & 0.0240 & 0.0173 & 50.0889$^\circ$ \\
12 & {\it Chili Pepper} & 0.0167 & 0.0446 & 0.0323 & 0.0306 & -0.0272 &  -86.4374$^\circ$ \\ 
13 & {\it Garlic} & 0.0100 & 0.0301 & 0.0293 & 0.0200 & -0.0147 & -57.6399$^\circ$ \\
14 & {\it Mushroom} & 0.0140 & 0.0545 & 0.0604 & 0.0342 & 0.0088 & 18.6744$^\circ$ \\
15 & {\it Watercress} & 0.0112 & 0.0658 & 0.0482 & 0.0385 & -0.0254 &  -69.0705$^\circ$ \\
16 & {\it Lentils} & 0.0095 & 0.0713 & 0.0338 & 0.0404 & 0.0252 & 104.7126$^\circ$ \\
17 & {\it Green Pepper} & 0.0324 & 0.0788 & 0.0506 & 0.0556 & -0.0503 & -95.6518$^\circ$ \\
18 & {\it Yam} & 0.0533 & 0.0724 & 0.0541 & 0.0628 & 0.0615 & 98.0833$^\circ$ \\
19 & {\it Tomato} & 0.0881 & 0.0679 & 0.0688 & 0.0780 & 0.0768 & 100.7557$^\circ$ \\
20 & {\it Pumpkin} & 0.0797 & 0.0713 & 0.0579 & 0.0755 & -0.0733 & -103.4804$^\circ$  \\
21 & {\it Broccoli} & 0.0143 & 0.1284 & 0.0642 & 0.0713 & -0.0422 & -99.6048$^\circ$ \\
22 & {\it Rice} & 0.0140 & 0.0412 & 0.0248 & 0.0276 & -0.0238 & -96.6635$^\circ$ \\ 
23 & {\it Parsley} & 0.0155 & 0.0266 & 0.0308 & 0.0210 & -0.0178 & -61.1698$^\circ$ \\
24 & {\it Black Pepper} & 0.0127 & 0.0294 & 0.0222 & 0.0211 & 0.0193 & 86.6308$^\circ$ \\
\hline
\end{tabular}
\end{center}
\caption{Interference data for concepts {\it A=Fruits} and {\it B=Vegetables}. The probability of a person choosing one of the exemplars as an example of {\it Fruits} (and as an example of {\it Vegetables}, respectively), is given by $\mu(A)$ (and $\mu(B)$, respectively) for each of the exemplars. The probability of a person choosing one of the exemplars as an example of {\it Fruits or Vegetables} is $\mu(A\ {\rm or}\ B)$ for each of the exemplars. The classical probability would be given by ${\mu(A)+\mu(B) \over 2}$, and $\theta$ is the quantum phase angle provoking the quantum interference effect.}
\end{table}

It should be noted that the data in Table 1 were not collected by actually putting the three questions to a fixed number of persons but derived from standard psychological experiments performed by James Hampton to measure the typicality of the 24 exemplars in Table 1 with respect to the concepts {\it Fruits}, {\it Vegetables} and their disjunction `{\it Fruits or Vegetables}' \cite{hampton1988}. Hampton asked the subjects to `estimate the typicality of the different exemplars with respect to the concepts {\it Fruits}, {\it Vegetables} and {\it Fruits or Vegetables}'. However, this set of data is appropriate for our experiment since the estimated typicality of an exemplar is strongly correlated with the frequency with which it is chosen as `a good example'. Preference is given to measuring typicality by asking subjects to estimate it, because this approach requires a much smaller sample of subjects to find statistically relevant results -- 40 in the case of Hampton's experiment. The reason is that each subject involved in an estimation test gives 24 answers, while in a choice test they give only one. This difference is irrelevant to our proposal for a new interpretation of quantum mechanics. We refer to \cite{aerts2009b} for a detailed description of the calculation of the frequency data of Table 1 from Hampton's typicality data. In \cite{aerts2009a}, we construct explicitly and in detail the quantum mechanical model for the pair of concepts {\it Fruit} and {\it Vegetable} and their disjunction `{\it Fruit or Vegetable}', showing that quantum interference models the experimental results. Here we directly give the details of the quantum mechanical model for the data of Table 1. We prove in \cite{aerts2009a} that it is possible to find a representation in $\compl^{25}$, where $|A\rangle$, representing the state of concept $A$, {\it Fruits}, is given by (\ref{vectorA}), and $|B\rangle$, representing the state of concept $B$, {\it Vegetables}, by (\ref{vectorB}). We have $\langle A|B\rangle=0$ and hence the vector ${1 \over \sqrt{2}}(|A\rangle+|B\rangle)$ represents the state of concept `$A$ or $B$', `{\it Fruits or Vegetables}'.
\begin{eqnarray} \label{vectorA}
|A\rangle&=&(\sqrt{\mu(A)_1},\ldots,\sqrt{\mu(A)_m},\ldots,\sqrt{\mu(A)_{24}},0) \\ \label{vectorB}
|B\rangle&=&(e^{i\beta_1}\sqrt{\mu(B)_1},\ldots,c_me^{i\beta_m}\sqrt{\mu(B)_m},\ldots,e^{i\beta_{24}}\sqrt{\mu(B)_{24}},\sqrt{\mu(B)_m(1-c_m^2)}) \\ \label{anglebetam}
\beta_m&=&\arccos({2\mu(A\ {\rm or}\ B)_m-\mu(A)_m-\mu(B)_m \over 2c_m\sqrt{\mu(A)_m\mu(B)_m}}) \\
 \label{anglebetak}
\beta_k&=&\pm\arccos({2\mu(A\ {\rm or}\ B)_k-\mu(A)_k-\mu(B)_k \over 2\sqrt{\mu(A)_k\mu(B)_k}}) 
\end{eqnarray}
The number $0\le c_m\le1$ is given by
\begin{equation} \label{cmequation}
c_m=\sqrt{{(-\sum_{k\not=m}\lambda_k)^2+(\mu(A\ {\rm or}\ B)_m-{\mu(A)_m+\mu(B)_m \over 2})^2 \over \mu(A)_m\mu(B)_m}}
\end{equation}
where $\lambda_k$ is
\begin{equation} \label{lambdak}
\lambda_k=\pm\sqrt{\mu(A)_k\mu(B)_k-(\mu(A\ {\rm or}\ B)_k-{\mu(A)_k+\mu(B)_k \over 2})^2}
\end{equation}
We prove in \cite{aerts2009a} that it is always possible to choose the signs for $\lambda_k$ in (\ref{lambdak}) such that $0\le c_m\le1$ given in (\ref{cmequation}) is well defined, if $m\in\{1,\ldots,24\}$ is chosen such that $|\lambda_m|$ is the biggest of all $|\lambda_k|$ -- and hence this is how we choose $m$ -- and in correspondence with this choice of signs for $\lambda_k$ the signs for $\beta_k$ are chosen in (\ref{anglebetak}). We also determine an algorithm in \cite{aerts2009a} for these choices. The measurement `a good example of' is represented by means of a self-adjoint operator with spectral decomposition $\{M_k\ \vert\ k=1,\ldots,24\}$, where $M_k$ corresponds to item $k$ of the list of items in Table 1. For $k\not=m$ we choose $M_k$ to be the orthogonal projection on canonical base vector number $k$ of $\compl^{25}$, while $M_n$ is the orthogonal projection on the plane spanned by canonical base vector $m$ and canonical base vector $25$. Following the standard rules of quantum mechanics the probabilities $\mu(A)_k$, $\mu(B)_k$ and $\mu(A\ {\rm or}\ B)_k$ are given by $\mu(A)_k=\langle A|M_k|A\rangle$, $\mu(B)_k=\langle B|M_k|B\rangle$ and $\mu(A\ {\rm or}\ B)_k={1 \over 2}\langle A+B|M_k|A+B\rangle$, and a straightforward calculation gives
\begin{equation} \label{muAorB}
\mu(A\ {\rm or}\ B)_k={1 \over 2}(\langle A|M_k|A\rangle+\langle B|M_k|B\rangle+\langle A|M_k|B\rangle+\langle B|M_k|A\rangle)={1 \over 2}(\mu(A)_k+\mu(B)_k)+\Re\langle A|M_k|B\rangle
\end{equation}
where $\Re\langle A|M_k|B\rangle$ is the interference term.

Let us construct this quantum model for the data given in Table 1, i.e. the data collected in \cite{hampton1988}. The exemplar which gives rise to the biggest value of $|\lambda_k|$ is {\it Tomato}, and hence we choose $m=19$. The values of $\lambda_k$ as given in (\ref{lambdak}) with their signs determined using the algorithm in \cite{aerts2009a} are given in Table 1, and also the values of the quantum phases as given in (\ref{anglebetam}) and (\ref{anglebetak}) with corresponding signs are given in Table 1. For the vectors we find
\begin{eqnarray}
\!\!\!\!\!\!\!\! |A\rangle&=&(0.1895, 0.2061, 0.1929, 0.2421, 0.2748, 0.3204, 0.3373, 0.3441, 0.1222, 0.1165, 0.1252, 0.1291, \nonumber \\
&&0.1002, 0.1182, 0.1059, 0.0974, 0.1800, 0.2308, 0.2967, 0.2823, 0.1194, 0.1181, 0.1245, 0.1128, 0) \\
\!\!\!\!\!\!\!\! |B\rangle&=&(0.1154e^{i83.8854^\circ}, 0.1040e^{-i94.5520^\circ}, 0.1484e^{-i95.3620^\circ}, 0.1640e^{i91.8715^\circ}, 0.1120e^{i57.9533^\circ}, \nonumber \\
&&0.1302e^{i95.8648^\circ}, 0.1302e^{-i113.2431^\circ}, 0.1246e^{i87.6039^\circ}, 0.1580e^{-i105.9806^\circ}, 0.1596e^{i99.3810^\circ}, \nonumber \\
&&0.1798e^{i50.0889^\circ}, 0.2112e^{-i86.4374^\circ}, 0.1734e^{-i57.6399^\circ}, 0.2334e^{i18.6744^\circ}, 0.2565e^{-i69.0705^\circ},  \nonumber \\
&&0.2670e^{i104.7126^\circ}, 0.2806e^{-i95.6518^\circ}, 0.2690e^{i98.0833^\circ}, 0.2606e^{i100.7557^\circ}, 0.2670e^{-i103.4804^\circ},  \nonumber \\
&&0.3584e^{-i99.6048^\circ}, 0.2031e^{-i96.6635^\circ}, 0.1630e^{-i61.1698^\circ}, 0.1716e^{i86.6308^\circ}, 0.1565).
\end{eqnarray}
This proves that we can model the data of \cite{hampton1988} by means of a quantum mechanical model, and such that the values of $\mu(A\ {\rm or}\ B)_k$ are determined from the values of $\mu(A)_k$ and $\mu(B)_k$ as a consequence of quantum interference effects.

In \cite{aerts2007a,aerts2007b,aerts2009c} we analyzed the origin of the interference effects that are produced when concepts are combined, and we provided an initial explanation that we investigated further in \cite{aertsdhooghe2009}. We will now show that this explanation, in addition to helping to gain a better understanding of the meaning of our hypothesis {\it NQE}, provides a new and surprising clarification of the interference of quantum entities themselves. To make this clear we need to take a closer look at the experimental data and how they are produced by interference. The exemplars for which the interference is a weakening effect, i.e. where $\mu(A\ {\rm or}\ B) < 1/2(\mu(A)+\mu(B))$ or $90^\circ \le \theta$ or $\theta \le -90^\circ$, are the following (in decreasing order of weakening effect): {\it Elderberry}, {\it Mustard}, {\it Lentils}, {\it Pumpkin}, {\it Tomato}, {\it Broccoli}, {\it Wheat}, {\it Yam}, {\it Rice}, {\it Raisin}, {\it Green Pepper}, {\it Peanut}, {\it Acorn} and {\it Olive}. The exemplars for which interference is a strengthening effect, i.e. where $1/2(\mu(A)+\mu(B)) < \mu(A\ {\rm or}\ B)$ or $\theta < 90^\circ$ or $-90^\circ \le \theta$, are the following (in decreasing order of strengthening effect): {\it Mushroom}, {\it Root Ginger}, {\it Garlic}, {\it Coconut}, {\it Parsley}, {\it Almond}, {\it Chili Pepper}, {\it Black Pepper}, and {\it Apple}. Let us consider the two extreme cases, viz. {\it Elderberry}, for which interference is the most weakening ($\theta=-113.2431^\circ$), and {\it Mushroom}, for which it is the most strengthening ($\theta=18.6744$). For {\it Elderberry}, we have $\mu(A)=0.1138$ and $\mu(B)=0.0170$, which means that test subjects have classified {\it Elderberry} very strongly as {\it Fruits} ({\it Apple} is the most strongly classified as {\it Fruits}, but {\it Elderberry} is next and close to it), and quite weakly as {\it Vegetables}. For {\it Mushroom}, we have $\mu(A)=0.0140$ and $\mu(B)=0.0545$, which means that test subjects have weakly classified {\it Mushroom} as {\it Fruits} and moderately as {\it Vegetables}. Let us suppose that $1/2(\mu(A)+\mu(B))$ is the value estimated by test subjects for `{\it Fruits or Vegetables}'. In that case, the estimates for {\it Fruits} and {\it Vegetables} apart would be carried over in a determined way to the estimate for `{\it Fruits or Vegetables}', just by applying this formula. This is indeed what would be the case if the decision process taking place in the human mind worked as if a classical particle passing through the {\it Fruits} hole or through the {\it Vegetables} hole hit the mind and left a spot at the location of one of the exemplars. However, in reality the situation is more complicated. When a test subject makes an estimate with respect to `{\it Fruits or Vegetables}', a new concept emerges, namely the concept `{\it Fruits or Vegetables}'. For example, in answering the question whether the exemplar {\it Mushroom} is a good example of `{\it Fruits or Vegetables}', the subject will consider two aspects or contributions. The first is related to the estimation of whether {\it Mushroom} is a good example of {\it Fruits} and to the estimation of whether {\it Mushroom} is a good example of {\it Vegetables}, i.e. to estimates of each of the concepts separately. It is covered by the formula $1/2(\mu(A)+\mu(B))$. The second contribution concerns the test subject's estimate of whether or not {\it Mushroom} belongs to the category of exemplars that cannot readily be classified as {\it Fruits} or {\it Vegetables}. This is the class characterized by the newly emerged concept `{\it Fruits or Vegetables}'. And as we know, {\it Mushroom} is a typical case of an exemplar that is not easy to classify as `{\it Fruits or Vegetables}'. That is why {\it Mushroom}, although only slightly covered by the formula $1/2(\mu(A)+\mu(B))$, has an overall high score as `{\it Fruits or Vegetables}'. The effect of interference allows adding the extra value to $1/2(\mu(A)+\mu(B))$ resulting from the fact that {\it Mushroom} scores well as an exemplar that is not readily classified as `{\it Fruits or Vegetables}'. This explains why {\it Mushroom} receives a strengthening interference effect, which adds to the probability of it being chosen as a good example of `{\it Fruits or Vegetables}'. {\it Elderberry} shows the contrary. Formula $1/2(\mu(A)+\mu(B))$ produces a score that is too high compared to the experimentally tested value of the probability of its being chosen as a good example of `{\it Fruits or Vegetables}'. The interference effect corrects this, subtracting a value from $1/2(\mu(A)+\mu(B))$. This corresponds to the test subjects considering {\it Elderberry} `not at all' to belong to a category of exemplars hard to classify as {\it Fruits} or {\it Vegetables}, but rather the contrary. As a consequence, with respect to the newly emerged concept `{\it Fruits or Vegetables}', the exemplar {\it Elderberry} scores very low, and hence the $1/2(\mu(A)+\mu(B))$ needs to be corrected by subtracting the second contribution, the quantum interference term. A similar explanation of the interference of {\it Fruits} and {\it Vegetables} can be put forward for all the other exemplars. The following is a general presentation of this. `For two concepts $A$ and $B$, with probabilities $\mu(A)$ and $\mu(A)$ for an exemplar to be chosen as a good example of $A$ and $B$, respectively, the interference effect allows taking into account the specific probability contribution for this exemplar to be chosen as a good exemplar of the newly emerged concept `$A\ {\rm or}\ B$', adding or subtracting to the value $1/2(\mu(A)+\mu(B))$, which is the average of $\mu(A)$ and $\mu(B)$'.

The foregoing analysis shows that there is a very straightforward and transparent explanation for the interference effect of concepts. In the following, we will show that this explanation leads to a new understanding of the interference of quantum entities themselves. Also, a detailed analysis of this explanation concerning the double-slit situation for quantum entities allows to give a far more detailed explanation of hypothesis {\it NQE}. Consider a typical double-slit situation in quantum mechanics. Figure 1 below presents the interference patterns obtained with both holes open (`$A$ and $B$ open quantum') and only one hole open (`$A$ open $B$ closed' and `$B$ open $A$ closed'), respectively. \begin{figure}[H]
\centerline {\includegraphics[width=11cm]{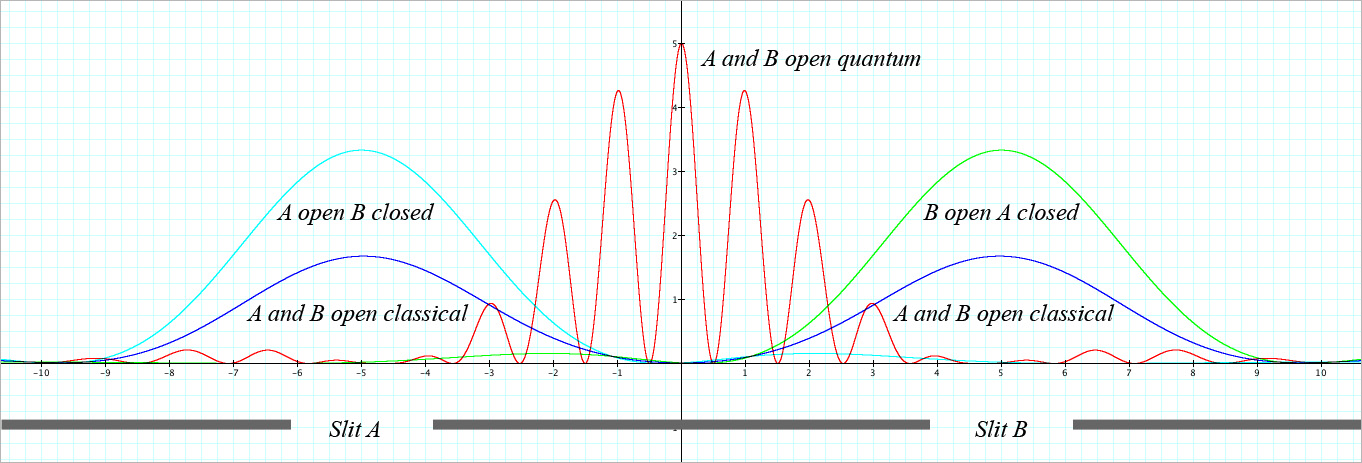}}
\caption{A typical interference pattern of a quantum two-slit situation with slits $A$ and $B$. The `{\it A open B closed}' curve represents the probability of detection of the quantum entity in case only {\it Slit A} is open; the `{\it B open A closed}' curve reflects the situation where only {\it Slit B} is open; and the `{\it A and B open classical}' curve is the average of both. The `{\it A and B open quantum}' curve represents the probability of detection of the quantum entity if both slits are open.}
\end{figure}
\noindent
Rather than presenting an image of a quantum entity passing through either or both slits `as an object would', we put forward a very different idea, namely the idea that the quantum entity passing through either or both slits `is' the conceptual entity standing for one of these situations. More concretely, we have a quantum entity, let us say `a photon'. This `is' a conceptual entity, hence `the photon is a photon as a concept'. This concept-photon can be in different states, and we will consider three of them: `State $A$ of the concept photon' is the conceptual combination: `the photon passes through hole $A$'.
`State $B$ of the concept photon' is the conceptual combination: `the photon passes through hole $B$'.
`State $A\ {\rm or}\ B$ of the concept photon' is the conceptual combination: `the photon passes through hole $A$ or passes through hole $B$' 

To recognize the analogy with our {\it Fruits} and {\it Vegetables} example, we need to consider how {\it Fruits} and {\it Vegetables} are two possible states of the concept {\it Food}. In this analogy, the conceptual combination `the photon passes through hole $A$' corresponds to the conceptual combination `this food item is a fruit', and the conceptual combination `the photon passes through hole $B$' corresponds to the conceptual combination `this food item is a vegetable'. The conceptual combination `the photon passes through hole $A$ or passes through hole $B$' corresponds to the conceptual combination `this food item is a fruit or is a vegetable'. The photon detected in a spot $X$ on the screen behind the holes is again a specific state of the concept-photon, corresponding to the conceptual combination `the photon is detected in spot $X$'. Compare this to how the different exemplars of Table 1 determine also states of {\it Food}, and hence also states of {\it Fruits} and states of {\it Vegetables}, `as concepts'. `Being detected in spot $X$' now corresponds with `spot $X$ being a good example'. Hence, instead of saying that `the photon passing through hole $A$ is detected in spot $X$', we should say `the photon in spot $X$ is a good example of the photon passing through hole $A$'. If we look at the typical interference pattern in Figure 1, we see that on the screen behind slits $A$ and $B$ we have almost zero probability for a photon to be detected in case both slits are open, while we have a very high probability for a photon to be detected on the screen in the center between both slits, completely contrary to what one would expect, if photons were objects flying through the slits and subsequently hitting the screen. 

Let us now analyze this experimental result according to our new interpretation. If both slits are open, this means that the photon is in the state of conceptual combination `the photon passes through slit $A$ `or' passes through slit $B$'. And indeed, for a photon hitting the screen in a spot exactly in between both slits, this would be the type of `state of the photon' raising most doubts as to whether it passed through slit $A$ or slit $B$. By contrast, photons appearing in the regions behind the slits -- in case both slits are open -- would not make us doubt as to the slit through which they have passed. On the contrary, we can be quite certain that photons showing behind a slit have come through that particular slit, so that this `is not a photon raising doubts about whether it has come through the one or through the other slit'. This means that `a photon in a spot $X$ in the center between both slits {\it is a good example} of a photon having passed through slit $A$ or having passed through slit $B$, whereas `a photon in a spot $X$ behind one of the slits {\it is not a good example} of a photon having passed through slit $A$ or having passed through slit $B$'.

The foregoing reasoning explains why the interference pattern of the photons as conceptual entities deviates from the average probabilities. Analogous to the {\it Fruits} and {\it Vegetables} example, this deviation is due to the fact that the `photon passing through slit $A$ or passing through slit $B$' is the conceptual disjunction of `the photon passing through slit $A$' and `the photon passing through slit $B$'. This means that it is completely natural for an estimate of the probabilities of choice with respect to `photon captured in spot $X$' for this conceptual disjunction, for example, to yield high estimates in the center between the two slits on the detection screen, since it is here that we will have `most doubts as to the slit through which the photon has passed'. Conversely, estimates will be low right behind both slits on the detection screen, since it is `these spots that leave little doubt as to the slit through which the photon has passed'.

We have presented the core elements of our explanation of quantum interference of quantum entities based on the new interpretation we introduced. For the sake of clarity, we have simplified our explanation. The conceptual combinations that we have been considering give indeed rise to more complicated effects than the one we analyzed in detail. We refer to \cite{aerts2009a,aerts2009b} for detailed analyses of such more complicated effects, which add detail to `how our new interpretation explains the interference of quantum entities'. However, the main idea is already contained in the simple case considered above, being that `the quantum entity behaves like a concept and not like an object'. In the next section we analyze entanglement and non-locality, one of the other fundamental problem situations of quantum theory.

\section{Entanglement and non-locality}

To illustrate how entanglement and non-locality appear in a natural way in the new interpretation we put forward here, we consider an example similar to the ones we analyzed in \cite{aertsczachordhooghe2006,aertsgabora2005a,aertsgabora2005b}. We regard the sentence `{\it The Animal Acts}' as a conceptual entity, hence as a combination of concepts, and, to test for the presence of entanglement, we verify the well-known Bell inequalities. For this, we consider two couples of exemplars or states of the concept {\it Animal}, namely {\it Horse}, {\it Bear} and {\it Tiger}, {\it Cat}, and also two couples of exemplars or states of the concept {\it Acts}, namely {\it Growls}, {\it Whinnies} and {\it Snorts}, {\it Meows}. Our first experiment $A$ consists in test subjects choosing between the two exemplars {\it Horse} and {\it Bear} to answer the question `is a good example of' the concept {\it Animal}, and we put $E(A)=+1$ if {\it Horse} is chosen, and hence the state of {\it Animal} changes to {\it Horse}, and $E(A)=-1$ if {\it Bear} is chosen, and hence the state of {\it Animal} changes to {\it Bear}, introducing in this way the function $E$ which measures the `expectation value' for the test outcomes concerned. Our second experiment $A'$ consists in test subjects choosing between the two exemplars {\it Tiger} and {\it Cat} to answer the question `is a good example of' the concept {\it Animal}, and we consistently put $E(A')=+1$ if {\it Tiger} is chosen and $E(A')=-1$ if {\it Cat} is chosen to introduce a measure of the expectation value. The third experiment $B$ consists in test subjects choosing between the two exemplars {\it Growls} and {\it Whinnies} to answer the question `is a good example of' the concept {\it Acts}, with $E(B)=+1$ if {\it Growls} is chosen and $E(B)=-1$ if {\it Whinnies} is chosen, and the fourth experiment $B'$ consists in test subjects choosing between the exemplars {\it Snorts} and {\it Meows} to anser the question `is a good example of' the concept {\it Acts}, with $E(B')=+1$ if {\it Snorts} is chosen and $E(B')=-1$ if {\it Meows} is chosen.

Now we consider coincidence experiments in combinations $AB$, $A'B$, $AB'$ and $A'B'$ for the conceptual combination {\it The Animal Acts}. Concretely, this means that, for example, test subjects taking part in the experiment $AB$, to answer the question `is a good example of', will choose between the four possibilities (1) {\it The Horse Growls}, (2) {\it The Bear Whinnies} -- and if one of these is chosen we put $E(AB)=+1$ -- and (3) {\it The Horse Whinnies}, (4) {\it The Bear Growls} -- and if one of these is chosen we put $E(AB)=-1$. For the coincidence experiment, $A'B$ subjects, to answer the question `is a good example of', will choose between (1) {\it The Tiger Growls}, (2) {\it The Cat Whinnies} -- and in case one of these is chosen we put $E(A'B)=+1$ -- and (3) {\it The Tiger Whinnies}, (4) {\it The Cat Growls} -- and in case one of these is chosen we put $E(A'B)=-1$. For the coincidence experiment, $AB'$ subjects, to answer the question `is a good example of', choose between (1) {\it The Horse Snorts}, (2) {\it The Bear Meows} -- and in case one of these is chosen we put $E(AB')=+1$ -- and (3) {\it The Horse Meows}, (4) {\it The Bear Snorts} -- and in case one of these is chosen we put $E(AB')=-1$. And finally, for the coincidence experiment, $A'B'$ subjects, to answer the question `is a good example of', will choose between (1) {\it The Tiger Snorts}, (2) {\it The Cat Meows} -- and in case one of these is chosen we put $E(A'B')=+1$ -- and (3) {\it The Tiger Meows}, (4) {\it The Cat Snorts} -- and in case one of these is chosen we put $E(A'B')=-1$.
Quite obviously, in coincidence experiment $AB$, both {\it The Horse Whinnies} and {\it The Bear Growls} will yield rather high scores, with the two remaining possibilities {\it The Horse Growls} and {\it The Bear Whinnies} being chosen little. This means that we will get $E(AB)$ close to -1. On the other hand, in the coincidence experiment $A'B$ one of the four choices will be prominent, namely {\it The Tiger Growls}, while the three other possibilities, {\it The Cat Whinnies}, {\it The Tiger Whinnies}, and {\it The Cat Growls}, will be much less present amongst the choices made by the test subjects. This means that we have $E(A'B)$ close to +1. In the two remaining coincidence experiments, we equally have that only one of the choices is prominent. For $AB'$, this is {\it The Horse Snorts}, with the other three {\it The Bear Meows}, {\it The Horse Meows} and {\it The Bear Snorts} being much less present. For $A'B'$, the prominent choice is {\it The Cat Meows}, while the other three {\it The Tiger Snorts}, {\it The Tiger Meows} and {\it The Cat Snorts} are much less present. This means that we have $E(AB')$ is close to +1 and $E(A'B')$ is close to +1. If we now substitute the different expectation values related to the coincidence experiments in the Clauser-Horne-Shimony-Holt variant of Bell inequalities \cite{clauserhorneshimonyholt1969}, we get $E(A'B')+E(A'B)+E(AB')-E(AB)$ bigger than +2, taking into account that we would even get +4 in case -1, +1, +1 and +1 were the outcome for each one of the members of the expression. Since the Clauser-Horne-Shimony-Holt variant of Bell inequalities consists in requiring this expression to be contained in the interval $[-2,+2]$, our calculation shows that our example constitutes an explicit violation of the Bell inequalities.

We do not doubt that the Bell inequalities will be violated when data are collected from an experiment with test subjects. However, instead of performing such an experiment and using the collected data, we present another way to collect relevant data for this situation, namely by making use of the World Wide Web. We do this because it sheds light on an interesting aspect of our new quantum interpretation. As a consequence, it also reveals new aspects of the violation of the Bell inequalities by quantum entities. We first explain what we do and then analyze. We use Google, and search the numbers of pages that contain the different combinations of items as they appear in our example. More concretely, for the coincidence experiment $AB$, we find $670$ pages that contain the sentence `horse growls', $5$ pages that contain `bear whinnies', $5,650$ pages that contain `horse whinnies' and $44,800$ pages that contain `bear growls'. This means that on a totality of $670+5+5,650+44,800=51,125$ pages we get fractions of $670$, $5$, $5,650$ and $44,800$ for the different combinations considered. We suppose that each page is elected with equal probability, which allows us to calculate the probability for one of the combinations to be elected. This gives $P(A_1,B_1)=670/51,125=0.0131$ for `horse growls', $P(A_2,B_2)=5/51,125=9.78\times 10^{-05}$ for `bear whinnies', $P(A_1,B_2)=5,650/51,125=0.1105$ for `horse whinnies' and $P(A_2,B_1)=44,800/51,125=0.8763$ for `bear growls'. Knowing these probabilities, we can again calculate the expectation value for this coincidence experiment by means of the equation $E(A,B)=P(A_1,B_1)+P(A_2,B_2)-P(A_2,B_1)-P(A_1,B_2)=-0.9736$. We calculate the expectation values $E(A',B)$, $E(A,B')$ and $E(A',B')$ in an analogous way, making use of the Google results. For the coincidence experiment $A'B$, we find $5,500$ pages that contain `tiger growls', $4$ pages with `cat whinnies', $0$ pages with `tiger whinnies' and $2,530$ pages with `cat growls'. This gives $P(A'_1,B_1)=0.6846$, $P(A'_2,B_2)=0.0005$, $P(A'_1, B_2)=0$ and $P(A'_2,B_1)=0.3149$ and $E(A',B)=0.3702$. For the coincidence experiment $AB'$, we find $11,900$ pages that contain `horse snorts', $156$ pages with `bear meows', $41$ pages with `horse meows' and $897$ pages with `bear snorts'. This gives $P(A_1,B'_1)=0.9158$, $P(A_2,B'_2)=0.0120$, $P(A_1, B'_2)=0.0032$ and $P(A_2,B'_1)=0.0690$ and $E(A,B')=0.8556$. For the coincidence experiment $A'B'$, we find $96$ pages that contain `tiger snorts', $26,500$ pages with `cat meows', $163$ pages with `tiger meows' and $5,040$ pages with `cat snorts'. This gives $P(A'_1,B'_1)=0.0030$, $P(A'_2,B'_2)=0.8334$, $P(A'_1, B'_2)=0.0051$ and $P(A'_2,B'_1)=0.1585$ and $E(A',B')=0.6728$. For the expression appearing in the Clauser-Horne-Shimony-Holt variant of Bell inequalities, we get
\begin{equation}
E(A'B')+E(A'B)+E(AB')-E(AB)=2.8722
\end{equation}
which is manifestly greater than 2, and hence constitutes a violation of Bell inequalities. We did not take into account that in general there will be webpages containing several of the expressions considered. This means that the probabilities for a page to contain one of the expressions will be different from what we calculated here. However, in Google, and other search engines we considered, it is not possible to obtain the information about any such pages that contain more than one of the expressions. In \cite{aerts2009a}, we analyze this problem in detail and show that it does not affect the core of the violation of the Bell inequalities that we consider here. Since data collected in Google in this way change slightly over time, because new webpages are added to the Google database constantly, we mention that the data we used were collected on July 30, 2009.

Before presenting our analysis, we will discuss another situation for which we consider Bell inequalities. We suppose that there are two separated sources of knowledge. Since we have no two separated World Wide Webs available, we use for both sources the one World Wide Web, and will see that this is no problem for what we want to show. Consider one of the coincidence experiments, for example $AB$. We now choose one page from one of the sources of knowledge and suppose that this page contains the word `horse', and in parallel to this we choose one page from the second source of knowledge and suppose that this page contains the word `growls'. The joint page of these two pages is considered as a page that counts for the conceptual combination {\it The Horse Growls}.
Again, we can calculate the probabilities and expectation values. However, this time we have to proceed as follows. We search in Google and find 169,000,000 pages containing `horse' and 176,000,000 pages containing `bear'. This means that the probability for a page coming from the first source of knowledge to contain `horse' is given by $P(A_1)=169,000,000/(169,000,000+176,000,000)=0.4899$, and the probability for such a page to contain `bear' is given by $P(A_2)=176,000,000/(169,000,000+176,000,000)=0.5101$. Analogously, the probability for a page of the second source of knowledge to contain `growls' is $P(B_1)=1,420,000/(1,420,000+60,800)=0.9589$, since 1,420,000 is the number of pages found in Google that contain `growls', and the probability for a page of the second source of knowledge to contain `whinnies' is $P(B_2)=60,800/(1,420,000+60,800)=0.0412$, since 60,800 is the number of pages that contain `whinnies'. Since a page contains the pair `horse' and `growls' if it is the joint page of one page containing `horse' coming from the first source of knowledge and a second page containing `growls' and coming from the second source of knowledge, it follows that the probability for this to take place is $P(A_1,B_1)=P(A_1)P(B_1)=0.4697$. Analogously we find $P(A_2,B_2)=P(A_2)P(B_2)=0.0209$, $P(A_1,B_2)=P(A_1)P(B_2)=0.0201$ and $P(A_2,B_1)=P(A_2)P(B_1)=0.4892$. This gives $E(A,B)=-0.0186$. We calculate $E(A',B)$, $E(A,B')$ and $E(A',B')$ in an analogous way. The number of pages containing `tiger' is 107,000,000, the number of pages containing `cat' is 721,000,000, the number of pages containing `snorts' is 449,000 and the number of pages containing `meows' is 349,000. This gives $P(A'_1)=0.1292$, $P(A'_2)=0.8708$, $P(B'_1)=0.5627$ and $P(B'_2)=0.4373$. From this it follows that $P(A'_1,B_1)=P(A'_1)P(B_1)=0.1239$, $P(A'_2,B_2)=P(A'_2)P(B_2)=0.0358$, $P(A'_1,B_2)=P(A'_1)P(B_2)=0.0053$ and $P(A'_2,B_1)=P(A'_2)P(B_1)=0.8350$, and as a consequence we have $E(A'B)=-0.6807$. We get in an analogous way $E(A,B')=-0.0025$ and $E(A',B')=-0.0929$. For the expression appearing in the Clauser-Horne-Shimony-Holt variant of Bell inequalities this gives
\begin{equation}
E(A'B')+E(A'B)+E(AB')-E(AB)=-0.7575
\end{equation}
which is very different from the earlier obtained expression, and also does not violate Bell inequalities. Remark that again we did not take into account that a substantial part of he webpages contains the two expressions considered, but, as we analyze more in detail in \cite{aerts2009a}, this does not influence our result. The reason for this is that for the situation of `separated sources of knowledge', the non-violation of the Bell inequalities is structural, which means that it will not be affected by this aspect. Let us show this by first proving the following lemma: {
If $x$, $x'$, $y$ and $y'$ are real numbers such that $-1\le x, x', y , y\le +1$ and $S=xy + xy' + x'y - x'y'$ then $-2\le S \le +2$.} Proof: Since $S$ is linear in all four variables $x$, $x'$, $y$, $y'$, it must take on its maximum and minimum values at the corners of the domain of this quadruple of variables, that is, where each of $x$, $x'$, $y$, $y'$ is +1 or -1. Hence at these corners $S$ can only be an integer between -4 and +4. But $S$ can be rewritten as $(x + x')(y + y') - 2x'y'$, and the two quantities in parentheses can only be 0, 2, or -2, while the last term can only be -2 or +2, so that S cannot equal -3, +3, -4, or +4 at the corners. 

Since in the situation considered we have $P(A_i,B_j)=P(A_i)P(B_j)$, $P(A'_i,B_j)=P(A'_i)P(B_j)$, $P(A_i,B'_j)=P(A_i)P(B'_j)$ and $P(A'_i,B'_j)=P(A'_i)P(B'_j)$ we have $E(A,B)=E(A)E(B)$, $E(A',B)=E(A')E(B)$, $E(A,B')=E(A)E(B')$ and $E(A',B')=E(A')E(B')$, and hence from the lemma it follows that $-2\le E(A'B')+E(A'B)+E(AB')-E(AB) \le +2$, which proves the Clauser-Horne-Shimony-Holt variant of Bell inequalities to be valid.

The foregoing examples and analyses show that the crucial matter in the violation of Bell inequalities is the non-product nature of the probabilities $P(A_i,B_j)$, $P(A'_i,B_j)$, $P(A_i,B'_j)$ and $P(A'_i,B'_j)$, hence more specifically that for example $P(A_i,B_j)\not=P(A_i)P(B_j)$. If we understand why these coincidence probabilities are not of the product nature we can get an insight into the origin of the violation of Bell inequalities for the situations that we consider. The reason is simple. Consider for example $P(A_1,B_1)$ and let us analyze why it is different from $P(A_1)P(B_1)$. We have that $P(A_1,B_1)$ is the probability that a page chosen at random of the World Wide Web contains the sentence part `horse growls' in our first example, and then we find $P(A_1,B_1)=0.0146$. While $P(A_1)P(B_1)$ is the probability that for two pages chosen at random, one contains `horse' and the other contains `growls', and then we find $P(A_2)P(B_2)=0.4899$. These values are very different and it is easy to understand why. The probability to find the sentence part `horse growls' is little, because any meaning this sentence may have will not be easily ascertained, for the simple reason that it is most unusual for horses to growl. If however two `separated' or `independent' pages are chosen at random, the probability that `horse' appears on one of the pages, and `growls' on the other, is substantial. The fundamental reason for this difference is that in the second case the pages are `separated' or `independent', or rather, `not connected by meaning'. It is indeed because `one webpage' contains concepts that are all connected by meaning, that the occurrence of `horse growls' on one webpage is so small. We will further analyze this insight in the next section, where we make explicit what is the `micro realm' and what is the `macro realm' according to our new interpretation, because we want to finish this section by showing the way in which the general type of non locality that we are so used to being confronted with in quantum entanglement situations, appears in exactly the same way in the conceptual situations that we have considered.

So we want to show now that for concept combinations the violation of Bell inequalities -- and hence the presence of entanglement -- has exactly the same mathematical origin as for quantum entities. For quantum entities, entanglement is mathematically structurally due to the wave function $\psi(x,y)$ of a joint quantum entity of two quantum entities not necessarily being a product $\psi_1(x)\psi_2(y)$ of the wave function $\psi_1(x)$ of one of the quantum entities with the wave function $\psi_2(y)$ of the other quantum entity. Let us show that this is what also happens for concepts when they are combined. As said, human or artificial memory structures relate to concepts like macroscopic material entities -- such as measuring apparatuses -- relate to quantum particles. Macroscopic three-dimensional space is considered to be the theatre of the macroscopic material entities, i.e. the collection of `locations' where such macroscopic material entities can be situated, and also the medium through which quantum particles communicate with macroscopic material entities, treated as measurement apparatuses in the formalism of quantum mechanics.

The items {\it Horse}, {\it Bear} and {\it Tiger} are exemplars of the concept {\it Animal} and the items {\it Growls}, {\it Whinnies} and {\it Snorts} are exemplars of the concept {\it Acts}. There are many more exemplars of both concepts. Let us consider for instance {\it Dog}, {\it Cow}, etc\ldots and {\it Runs}, {\it Jumps}, etc \ldots. For the specific type of measurement that we have considered with respect to Bell inequalities -- `is a good example of' -- we can consider the different exemplars of both concepts as points in the memory structure of a human mind, and denote them $x_1, x_2, \ldots, x_n$ for {\it Animal} and $y_1, y_2, \ldots, y_m$ for {\it Acts}. {\it Animal} can then be written as a wave function $\psi_1(x)$, where $x$ can take the values $x_1, x_2, \ldots, x_n$, and $|\psi_1(x_i)|^2$ is the weight of the exemplar $x_i$ for the concept {\it Animal} with respect to the measurement `is a good example of {\it Animal}'. Likewise, the concept {\it Acts} can be represented by the wave function $\psi_2(y)$, where $y$ takes the values $y_1, y_2, \ldots, y_m$ and $|\psi(y_j)|^2$ is the weight of exemplar $y_j$ for the concept {\it Acts} with respect to the measurement `is a good example of {\it Acts}'. If the concepts {\it Animal} and {\it Acts} are combined to form a phrase, for example {\it The Animal Acts}, we can describe this by means of a wave function $\psi(x,y)$, where $|\psi(x_i,y_j)|^2$ is the weight of the couple of exemplars $x_i$ and $y_j$ for the considered combination of the concepts {\it Animal} and {\it Acts} with respect to the measurement `is a good example of {\it The Animal Acts}'. The foregoing analysis shows exactly that $\psi(x,y)$ is `not' a product. Indeed, if it was, Bell inequalities would not be violated. The reason `why it is not a product' is clearly illustrated by the above combination of {\it Animal} and {\it Acts} into {\it The Animal Acts}. This combination introduces a wave function that attributes weights to couples of exemplars $(x_i, y_j)$ in a new way, i.e. different from how weights are attributed by component wave functions describing the measurements related to `is this a good example of {\it Animal}' and `is this a good example of {\it Acts}' apart, and a product of such component wave functions. This is because {\it The Animal Acts} is not only a combination of concepts, but a new concept on its own account. It is this new concept that determines the values attributed to weights of couples of exemplars, which will therefore be different from the values attributed if we consider only the products of weights determined by the constituent concepts. Likewise, it is clear that `all functions of two variables', i.e. all functions $\psi(x,y)$, will be possible expressions of states of this new concept. Indeed, all types of combinations, e.g. {\it The Animal Acts with Courage}, {\it Look how this Animal Enjoys Swimming}, etc\ldots, will introduce different states $\psi(x,y)$ which are wave functions in the product space $\{x_1, \ldots, x_n\}\times\{y_1,\ldots,y_m\}$. This shows that combining concepts in a natural and understandable way gives rise to entanglement, and it does this structurally in a completely analogous way as entanglement appears in quantum mechanics, namely by allowing all functions of joint variables of two entities to play a role as wave functions describing states of the joint entity consisting of these two entities.

\section{Space, Time, Momentum, Energy, Prototypes and Evolution}
In \cite{aerts2009a}, in addition to entanglement and interference, we analyzed the aspects of quantum particles and concepts related to identity and individuality. Identical quantum particles behave in a very specific way, adding strongly to the problem of finding an explanatory framework for quantum physics, while in \cite{aerts2009a} we showed that concepts behave in a similar way, giving rise to Fermi Dirac and Bose Einstein statistics depending on whether the concepts are situated in memory or not. We also showed how a Heisenberg type uncertainty for concepts appears in a natural way, and more specifically as follows. The very concrete forms of a concept correspond to very localized states of a quantum particle and the very abstract forms of a concept correspond to states of a quantum particle very localized in momentum space. In \cite{aerts2009a}, we illustrated this analysis of identity, individuality and Heisenberg uncertainty by considering the specific conceptual environment of the World Wide Web, and such that single words are the most abstract forms and entire webpages the most concrete forms of each of the concepts appearing at least once in the text of these webpages. For example, the word `cat' is the most abstract form of the concept {\it Cat}, while each webpage containing the word `cat' at least once is a most concrete form of the concept {\it Cat}.

Before proceeding, we want to remark that if we compare the behavior of quantum entities to the behavior of human concepts, we do not expect to find an isomorphic structure for both. We believe that human concepts and their interactions are at a very primitive stage of development as compared to quantum entities and their interactions. This means that, although we expect to find connections with a profound explanatory potential with respect to fundamental aspects of both situations, i.e. human concepts and their interactions and quantum particles and their interactions, we also expect to find a much less crystallized and organized form for human concepts than for quantum particles. It is within this expectation that the analysis put forward in \cite{aerts2009a} and in the present article needs to be understood. This means that we regard the actual structure of the physical universe, space, time, momentum, energy and quantum particles interacting with ordinary matter as emergent from a much more primitive situation of interacting conceptual entities and their memories. Consequently, as part of the elaboration of our overall explanatory framework, one of the research aims must be to investigate which structural properties, laws and axioms may characterize a weakly organized conceptual structure, such as the one actually existing for the case of human concepts and memories, and which additional structural properties, laws and axioms could make it into a much more strongly organized conceptual structure, such as the one of the physical universe, space time, momentum energy and quantum particles interacting with ordinary matter. To investigate these axiomatic aspects of our explanatory framework and interpretation, we can rely partly on results we obtained in quantum axiomatics itself, where the formalism of quantum physics in a complex Hilbert space is reconstructed starting from a very general situation, namely the situation of an arbitrary entity and its properties where the properties are operationally defined by means of corresponding yes-non experiment testing these properties \cite{aerts2009d,aertsaerts2004,aerts2002,aertscolebundersvandervoordevansteirteghem1999,piron1990,aerts1983,aerts1982,aerts1981}. We also want to investigate in which way the older, more mathematically oriented approaches to axiomatization of quantum mechanics can be partly adapted to this new situation \cite{beltrametticassinelli1981,beltramettivanfraassen1980,foulisrandall1980,hooker1979,piron1976,jauch1968,varadarajan1968,piron1964,mittelstaedt1963,mackey1963,mackey1957,birkhoffvonneumann1936}. Quantum axiomatics is meant to model a physical object in a very general way, step by step arriving at the Hilbert space model of a quantum particle; the quantum axiomatic approach is partly valid for our situation, because the nature of the general physical object to start with is not a priori determined, which means that it can be what we have called a conceptual entity, or a concept. Indeed, concepts are linked to sets of characterizing properties in the same way as physical objects are, and since in the quantum axiomatic approach the steps to come to a Hilbert space representation are consistently taken by considering as basic mathematical object the collection of properties of this general physical object, the approach remains partly valid when the physical object is replaced by a conceptual entity, and the same set of conditions is applied to the collection of properties of this conceptual entity. We refer to \cite{aerts2009d,aertsaerts2004} for a general overview of quantum axiomatics following the so-called Geneva-Brussels approach, noting that quantum axiomatics is sometimes specified as quantum logic or quantum probability, depending on its focus, and hence aspects of the work of scientists in quantum logic and quantum probability are relevant for the axiomatics to be elaborated with respect to the interpretation and explanatory framework we propose in the present article and in \cite{aerts2009a}. We have started such an approach to the axiomatization of concepts using a quantum axiomatic approach in \cite{aertsgabora2005a}, but since it would be outside the scope of this article to elaborate on the formal and technical aspects, we will concentrate on some of the steps in such an axiomatization procedure, also because this will again shed more light on the new interpretation and explanatory framework that we introduce in the present article and in \cite{aerts2009a}. More concretely, we want to analyze the way in which non-locality may possibly have taken shape in the course of such an
 ongoing growth of structure and organization, from a very weak conceptual structure, such as the one of human concepts and memories, to a very strong conceptual structure, such as the one of the physical universe, space, time, momentum, energy and quantum particles interacting with ordinary matter.

One of the very direct ways to see the difference between a concept and an object is the following. For two concepts $A$ and $B$, we have that `$A$ or $B$' is again a concept. Obviously, in case $A$ and $B$ are objects, then `$A$ or $B$' is not an object. More concretely, `$Apple$ or $Coconut$' is a concept in case $Apple$ and $Coconut$ are concepts, but `$Apple$ or $Coconut$' is not an object in case $Apple$ and $Coconut$ are objects. If our basic hypothesis {\it NQE} is correct, this distinction goes to the root of the problems we have in understanding quantum physics. Consider for example a quantum particle and two states $A$ and $B$ of this particle that are well localized, meaning that they are described by wave functions $\psi_A(x,y,z)$ and $\psi_B(x,y,z)$ which are localized wave packets, meaning that the square of their absolute values are Gaussians with a small standard deviation, and suppose that both wave packets $\psi_A(x,y,z)$ and $\psi_B(x,y,z)$ are centered around points $(x_A,y_A,z_A)$ and $(x_B,y_B,z_B)$, respectively, such that the distance $d((x_A,y_A,z_A),(x_B,y_B,z_B))$ between both points is large. The normalized superposition state of $\psi_A(x,y,z)$ and $\psi_B(x,y,z)$, namely ${1 \over \sqrt{2}}(\psi_A(x,y,z)+\psi_B(x,y,z))$, is then a very non-local state of the quantum particle. If the quantum particle is in this superposition state, it literally means that when a measurement of localization is performed, the particle is found with probability 1/2 close to $(x_A,y_A,z_A)$ and with probability 1/2 close to $(x_B,y_B,z_B)$, independently of how large $d((x_A,y_A,z_A),(x_B,y_B,z_B))$ is. Nowadays, these extreme non-local states are realized experimentally on a macroscopic scale \cite{tittelbrendelgisinherzogzbindengisin1998}, which means that they really exist in our ordinary everyday world. Let us remember that a `superposition' creates a state out of two other states such that this new state can collapse into these two other states, even if the two states are not at all close; and also that a superposition is what models the `or' with respect to human concepts \cite{aerts2009c}. When we use the notion `close' in space for localized state, we should use `similar' in a more general setting, such as the one of human concepts. And indeed, this is what the `or' connective does for concepts. If we consider concepts $A$ and $B$, then `$A$ or $B$' is a concept which can collapse into concept $A$ or into concept $B$, independently of whether concepts $A$ and $B$ are similar or not. Suppose $A$ is the concept {\it Furniture} and $B$ the concept {\it Bird}, then $A$ and $B$ are not at all similar concepts. However, the concept `{\it Furniture or Bird}' does exist as a concept, and it can collapse to one of its constituent concepts {\it Furniture} and {\it Bird}.

From this we can learn that if a structure of similarity is introduced into the set of concepts, the `or' connective will introduce the generalized effect of non-locality in many cases, i.e. when it connects concepts that are not similar to each other within this structure of similarity. This very general notion of non-locality defined with respect to a similarity relation can evolve step by step into a more specific one if additional structure is introduced on the states of the concepts and the set of their features. Let us consider as an example prototype theory, where the hypothesis is introduced that concepts are organized with respect to a prototype in a graded way \cite{rosch1973}. Hence the notion of distance for a particular exemplar of a concept with respect to the prototype of this concept can be introduced, giving rise to a metric structure. For the concept {\it Furniture}, the exemplar {\it Chair} would be closest to the prototype of {\it Furniture}, because as experiments indeed reveal, {\it Chair} is found to be the most typical type of furniture, and for the concept {\it Bird} the exemplar {\it Robin} would be closest to the prototype of {\it Bird} for the same reason \cite{rosch1975}. A concept such as `{\it Furniture or Bird}', however, is a concept for which it would be very difficult to define a prototype. Such concepts for which no obvious prototype exists are the equivalents of what in quantum mechanics are extreme non-local states of quantum particles. The analogue we put forward here shows us that non-locality is a consequence of `the choice for an ordering related to a measure of similarity', because even with a very weak ordering, such as the one introduced by prototype theory for the set of human concepts, the predecessor of non-locality appears. 

We believe that the `space, time, momentum, energy and quantum particles interacting with ordinary matter situation' of the universe, which is often considered the theatre of reality, is emergent, and orders the behavior of quantum particles and their interaction with ordinary matter in a kind of `best possible way', exactly like prototype theory orders human concepts in a best possible way. The appearance of non-locality for `quantum particles interacting with ordinary matter' with respect to space time and the appearance of `concepts without prototype' for human concepts are consequences of the choice for this ordering. And parallel to this, the choice for a specific ordering has consequences for the growth and evolution of the concepts themselves. Indeed, although `{\it Furniture or Bird}' definitely exists as a concept, we will not encounter it as a human in any obvious way in our daily world. It is much more likely that we will encounter concepts such as `{\it Fruits or Vegetables}', which is a much less `non-local' combination by means of the connective `or', since {\it Fruits} and {\it Vegetables} are both good exemplars of the concept {\it Food}. To find a concept for which {\it Furniture} and {\it Bird} are both good exemplars, we almost have to raise the level of abstraction to the concept {\it Thing} or {\it Object}, or even, in a pure way, to {\it Concept}. Indeed, the choice and identification of the notion of exemplar is already inspired by an intuitive ordering of human concepts similar to the prototype ordering. {\it Exemplars} are concepts that `spiral' around the core of a concept for which they are exemplars. Hence, it is difficult to use the notion of exemplars for a concept without prototype. We believe that also for quantum particles the growth and evolution of the `particles interacting with ordinary matter' has been influenced by the ordering structures appearing within their conceptual environment. This is the reason that non-locality is not present in abundance in the world of `quantum particles interacting with ordinary matter', exactly like concepts without a prototype are not present in abundance in the set of `evolved human concepts interacting with memory structures'. But the potential of forming concepts without prototype has remained completely intact, which is also the same for the case of quantum particles, where the potential to form non-local states has remained completely intact. This potential is exploited in fine and clever ways in the quantum physics laboratories, which are currently producing non-local quantum states in many different ways.

By putting forward the hypothesis that quantum particles are conceptual entities, we may have given the impression of intending to develop a radically anthropomorphic view of what goes on in the micro-world, claiming that `what happens in our macro-world, i.e. people using concepts and their combinations to communicate', already took place in the micro-realm, i.e. `measuring apparatuses, and more generally entities made of ordinary matter, communicating with each other, where the words and sentences of their language are the quantum particles'. Although this is certainly a fascinating and possibly also a feasible form of metaphysics compatible with the explanatory framework that we put forward, it is not a necessary consequence of our basic hypothesis. Further detailed research will be required to gain an initial understanding of which aspects of such a drastic metaphysical view are true and which are not at all. If instead of `conceptual entity' we use the notion of `sign', we can formulate our basic hypothesis in much less anthropomorphic terms as follows: `Quantum entities are signs exchanged between measuring apparatuses and more generally between entities made of ordinary matter'. We use the notion of `sign' here as it appears in semiotics \cite{chandler2002}, which is the study of the exchange of signs of any type, which means that it covers animal communication, but also the exchange of signs, including icons, between computer interfaces. This much less anthropomorphic view allows to interpret measurement apparatuses and more general entities made of ordinary matter as interfaces for these signs instead of memories for conceptual entities. Our basic hypothesis can be made even more general in the following way. The exchange between entities of ordinary matter with quantum particles as mediators can be a proto-situation of conceptual exchange between memory structures with concepts as mediators and/or semiotic exchange between interfaces with signs as mediators. This is similar to how the wave phenomenon is a proto-situation for electromagnetic waves as well as for sound waves and any other types of waves. However, what is a fundamental consequence of our basic hypothesis, whether we consider its `cognitive version', its `semiotic version' or its `proto-situation' version, is that communication of some type or at least mediation takes place, and, more specifically, that the language, system of signs or mediating entities used in this communication or mediation evolved symbiotically with the memories for this language, the interfaces for these signs or the entities mediated in between. This introduces in any case a radically new way to look upon the evolution of the part of the universe we live in, namely the part of the universe consisting of entities of ordinary matter and quantum particles. Any mechanistic view, whether the mechanistic entities are conceived of as particles, as waves or as both, cannot work out well if the reality is one of co-evolving concepts and memories, signs and interfaces or mediating entities and mediated in-between entities.


\small
\begin{thebibliography}{99} 
\setlength{\itemsep}{-1.5mm}
\bibitem{aerts2009a} Aerts, D.: Quantum particles as conceptual entities. A possible explanatory framework for quantum theory. {\it Foundations of Science}, {\bf 14}, 361-411 (2009). 

\bibitem{aerts2009b} Aerts, D.: Quantum interference in cognition and double-slit graphical representations. In preparation.


\bibitem{aerts2009c} Aerts, D.: Quantum structure in cognition. {\it Journal of Mathematical Psychology}, {\bf 53}, 314-348 (2009).

\bibitem{aertsaertsgabora2009} Aerts, D., Aerts, S. and Gabora, L.: Experimental evidence for quantum structure in cognition. {\it Lecture Notes in Artificial Intelligence}, {\bf 5494}, 59-70 (2009).

\bibitem{aertsdhooghe2009} Aerts, D. and D'Hooghe, B.: Classical logical versus quantum conceptual thought: Examples in economy, decision theory and concept theory. {\it Lecture Notes in Artificial Intelligence}, {\bf 5494}, 128-142 (2009).


\bibitem{aerts2007a} Aerts, D.: Quantum interference and superposition in cognition: Development of a theory for the disjunction of concepts. In D. Aerts, B. D'Hooghe and N. Note (Eds.), {\it Worldviews, Science and Us: Bridging Knowledge and Its Implications for Our Perspectives of the World}. Singapore: World Scientific (2010).

\bibitem{aerts2007b} Aerts, D.: General quantum modeling of combining concepts: A quantum field model in Fock space. In D. Aerts, B. D'Hooghe and N. Note (Eds.), {\it Worldviews, Science and Us: Bridging Knowledge and Its Implications for Our Perspectives of the World}. Singapore: World Scientific (2010). 

\bibitem{aertsczachordhooghe2006} Aerts, D., Czachor, M. and D'Hooghe, B.: Towards a quantum evolutionary scheme: violating Bell's inequalities in language. {\it Evolutionary Epistemology, Language and Culture -- A Non Adaptationist Systems Theoretical Approach.} Dordrecht: Springer (2006).

\bibitem{aertsgabora2005a} Aerts, D. and Gabora, L.: A theory of concepts and their combinations I: The structure of the sets of contexts and properties. {\it Kybernetes}, {\bf 34}, 167-191 (2005).

\bibitem{aertsgabora2005b} Aerts, D. and Gabora, L.: A theory of concepts and their combinations II: A Hilbert space representation. {\it Kybernetes}, {\bf 34}, 192-221 (2005).

\bibitem{aertsczachor2004} Aerts, D. and Czachor, M.: Quantum aspects of semantic analysis and symbolic artificial intelligence. {\it Journal of Physics A-Mathematical and General}, {\bf 37}, L123-L32 (2004). 

\bibitem{gaboraaerts2002} Gabora, L. and Aerts, D.: Contextualizing concepts using a mathematical generalization of the quantum formalism. {\it Journal of Experimental and Theoretical Artificial Intelligence}, {\bf 14}, 327-358 (2002).

\bibitem{aertsbroekaertsmets1999a} Aerts, D., Broekaert, J., Smets, S.: A quantum structure description of the liar paradox. {\it International Journal of Theoretical Physics}, {\bf 38}, 3231-3239 (1999).

\bibitem{aertsbroekaertsmets1999b} Aerts, D., Broekaert, J. and Smets, S.: The liar paradox in a quantum mechanical perspective. {\it Foundations of Science}, {\bf 4}, 115-132 (1999).

\bibitem{aertsaerts1994} Aerts, D., Aerts, S.: Applications of quantum statistics in psychological studies of decision processes. {\it Foundations of Science}, {\bf 1}, 85-97 (1994).

\bibitem{aertsaertsbroekaertgabora2000} Aerts, D., Aerts, S., Broekaert, J. and Gabora, L.: The violation of Bell inequalities in the macroworld. {\it Foundations of Physics}, {\bf 30}, 1387-1414 (2000). 

\bibitem{aerts1999a} Aerts, D.: The stuff the world is made of: physics and reality. {\it Einstein meets Magritte: An Interdisciplinary Reflection}. Dordrecht: Kluwer Academic (1999).

\bibitem{aerts1999b} Aerts, D.: Foundations of quantum physics: a general realistic and operational approach. {\it International Journal of Theoretical Physics}, {\bf 38}, 289-358 (1999). 

\bibitem{aerts1998a} Aerts, D.: The entity and modern physics: the creation-discovery view of reality. {\it Interpreting Bodies: Classical and Quantum Objects in Modern Physics}. Princeton: Princeton University Press (1998).

\bibitem{aerts1998b} Aerts, D.: The hidden measurement formalism: what can be explained and where paradoxes remain. {\it International Journal of Theoretical Physics}, {\bf 37}, 291-304 (1998).
 
\bibitem{aerts1995} Aerts, D.: Quantum structures: an attempt to explain their appearance in nature. {\it International Journal of Theoretical Physics}, {\bf 34}, 1165-1186 (1995). 

\bibitem{aerts1994} Aerts, D.: Quantum structures, separated physical entities and probability. {\it Foundations of Physics}, {\bf 24}, 1227-1259 (1994).

\bibitem{aerts1993} Aerts, D.: Quantum structures due to fluctuations of the measurement situations. {\it International Journal of Theoretical Physics}, {\bf 32}, 2207-2220 (1993).

\bibitem{aerts1992} Aerts, D.: The construction of reality and its influence on the understanding of quantum structures. {\it International Journal of Theoretical Physics}, {\bf 31}, 1815-1837 (1992).

\bibitem{aerts1986} Aerts, D.: A possible explanation for the probabilities of quantum mechanics. {\it Journal of Mathematical Physics}, {\bf 27}, 202-210 (1986).
\bibitem{tonomuraendomatsudakawasakiezawa1989} Tonomura, A., Endo, J., Matsuda, T., Kawasaki, T. and Ezawa, H.: Demonstration of single-electron buildup of an interference pattern, {\it American Journal of Physics}, {\bf 57}, 117-120 (1989).

\bibitem{jonsson1961} J\"onsson, C.: Elektroneninterferenzen an mehreren k\"unstlich hergestellten Feinspalten, {\it Zeitschrift f\"ur Physik}, {\bf 161}, 454-474, (1961).

\bibitem{young1802} Young, T.: On the theory of light and colours. {\it Philosophical Transactions of the Royal Society}, {\bf 92}, 12-48 (1802). Reprinted in part in H. Crew, (Ed.) (1990). {\it The Wave Theory of Light}. New York.

\bibitem{feynman1970} Feynman, R.: {\it The Feynman Lectures on Physics: Volume 3}. Reading Massachusetts: Addison Wesley Publishing Company (1970).

\bibitem{feynman1965} Feynman, R.: {\it The Character of Physical Law}. Massachusetts: The MIT Press (1965).

\bibitem{pearle2007} Pearle, Ph.: How stands collapse I. {\it Journal of Physics A: Mathematical and Theoretical}, {\bf 40}, 3189-3204 (2007).

\bibitem{gisin1989} Gisin, N.: Stochastic quantum dynamics and relativity. {\it Helvetica Physica Acta}, {\bf 62}, 363-371 (1989). 

\bibitem{ghirardiriminiweber1986} Ghirardi, G. C., Rimini, A. and Weber, T.: Unified dynamics for microscopic and macroscopic systems. {\it Physical Review D}, {\bf 34}, 470-491 (1986).

\bibitem{zbindenbrendeltittelgisin2001} Zbinden, H., Brendel, J., Tittel, W. and Gisin, N.: Experimental test of relativistic quantum state collapse with moving reference frames. {\it Journal of Physics A: Mathematical and Theoretical}, {\bf 34}, 7103-7109 (2001).

\bibitem{simonbuzekgisin2001} Simon, C., Buzek, V. and Gisin, N.: No-signaling condition and quantum dynamics. {\it Physical Review Letters}, {\bf 87}, 170405-2 (2001).

\bibitem{svetlichny1998} Svetlichny, G.: Quantum formalism with state-collapse and superluminal communication. {\it Foundations of Physics}, {\bf 28}, 131-155 (1998).

\bibitem{deutsch1999} Deutsch, D.: Quantum theory of probability and decisions. {\it Proceedings of the Royal Society of London}, {\bf A455}, 3129-3137 (1999).

\bibitem{dewittgraham1973} DeWitt, B. S., and Graham, N. (eds.): {\it The Many-Worlds Interpretation of Quantum Mechanics}. Princeton: Princeton University Press (1973).

\bibitem{everett1957} Everett, H.: Relative state formulation of quantum mechanics. {\it Reviews of Modern Physics}, {\bf 29} 454-462 (1957). 

\bibitem{aspectgrangierrobert1981} Aspect, A., Grangier, P. and Roger, G.: Experimental tests of realistic local theories via Bell's theorem. {\it Physical Review Letters}, {\bf 47}, 460-463 (1981).

\bibitem{bell1964} Bell, J. S.: On the Einstein-Podolsky-Rosen paradox. {\it Physics}, {\bf 1}, 195-200 (1964).

\bibitem{einsteinpodolskyrosen1935} Einstein, A., Podolsky, B. and Rosen, N.: Can quantum-mechanical description of physical reality be considered complete? {\it Physical Review}, {\bf 47}, 777-780 (1935).

\bibitem{salartbaasbranciardgisinzbinden2008} Salart, D., Baas, A., Branciard, C., Gisin, N. Zbinden, H.: Testing the speed of `spooky action at a distance. {\it Nature}, {\bf 454}, 861-864 (2008).

\bibitem{bleingmichaeljordan2003} Blei, D. M., Ng, A. N. and Jordan, Michael I.: Latent Dirichlet Allocation. {\it Journal of Machine Learning Research}, {\bf 3}, 993-1022 (2003).

\bibitem{griffithssteyvers2002} Griffiths, T. L. and Steyvers, M.: Prediction and semantic 
association. {\it Advances in Neural Information Processing Systems}, {\bf 15}, 11-18 (2002).

\bibitem{hofmann1999} Hofmann, T.: Probabilistic Latent Semantic Analysis. {\it Proceedings of the 22nd annual international ACM SIGIR conference on Research and development in information retrieval}, Berkeley, California, 50-57 (1999).

\bibitem{lundburgess1995} Lund, K. and Burgess, C.: Producing high-dimensional semantic spaces from lexical co-occurrence. {\it Behavior Research Methods, Instruments and Computers}, {\bf 28}, 203-208 (1995).

\bibitem{deerwesterdumaisfurnaslandauerharshman1990}Deerwester, S., Dumais, S. T., Furnas, G. W., Landauer, T. K. and Harshman, R.: Indexing by Latent Semantic Analysis. {\it Journal of the American Society for Information Science}, {\bf 41}, 391-407 (1990).

\bibitem{hampton1988} Hampton, J. A.: Disjunction of natural concepts.
{\it Memory \& Cognition}, {\bf 16}, 579-591 (1988).

\bibitem{clauserhorneshimonyholt1969} Clauser, J. F., Horne, M. A., Shimony, A., and Holt, R. A.: Proposed experiment to test local hidden-variable theories. {\it Physical Review Letters}, {\bf 23}, 880-884 (1969).

\bibitem{aerts2009d} Aerts, D.: Quantum axiomatics. In K. Engesser, D. Gabbay and D. Lehmann (Eds.), {\it Handbook of Quantum Logic, Quantum Structure and Quantum Computation}. Amsterdam: Elsevier (2009).

\bibitem{aertsaerts2004} Aerts, D. and Aerts. S.: Towards a general operational and realistic framework for quantum mechanics and relativity theory. In A. C. Elitzur, S. Dolev and N. Kolenda (Eds.), {\it Quo Vadis Quantum Mechanics? Possible Developments in Quantum Theory in the 21st Century} (pp. 153-208). New York: Springer (2004).

\bibitem{aerts2002} Aerts, D.: Being and change: foundations of a realistic operational formalism. In D. Aerts, M. Czachor and T. Durt, (Eds.), {\it Probing the 
Structure of Quantum Mechanics: Nonlinearity, Nonlocality, Probability and Axiomatics}. Singapore: World Scientific (2002).

\bibitem{aertscolebundersvandervoordevansteirteghem1999} Aerts, D., Colebunders, E., Van der Voorde, A. and Van Steirteghem, B.: State property systems and closure spaces: a study of categorical equivalence. {\it International Journal of Theoretical Physics}, {\bf 38}, 359-385 (1999).

\bibitem{piron1990} Piron, C.: {\it M\'ecanique Quantique: Bases et Applications}. Lausanne: Press Polytechnique de Lausanne (1990).

\bibitem{aerts1983} Aerts, D.: Classical theories and non classical theories as a special case of a more general theory. {\it Journal of Mathematical Physics}, {\bf 24}, 2441-2453 (1983).

\bibitem{aerts1982} Aerts, D.: Description of many physical entities without the paradoxes encountered in quantum mechanics. {\it Foundations of Physics}, {\bf 12}, 1131-1170 (1982).

\bibitem{aerts1981} Aerts, D.: {\it The One and the Many: Towards a Unification of the Quantum and Classical Description of One and Many Physical Entities}. Doctoral dissertation, Brussels Free University (1981).

\bibitem{beltrametticassinelli1981} Beltrametti, E. and Cassinelli, G.: The Logic of Quantum Mechanics. Reading, Massachusetts: Addison-Wesley (1981).

\bibitem{beltramettivanfraassen1980} Beltrametti, E. and van Fraassen, B. C. (Eds.): {\it Current Issues in Quantum Logic}. New York: Plenum (1980).

\bibitem{foulisrandall1980} Foulis, D. J., and Randall, C.H.: What are quantum logics and what ought they to be? In E. Beltrametti and B. C. van Fraassen, (Eds.), {\it Current Issues in Quantum Logic}. New York: Plenum (1980).

\bibitem{hooker1979} Hooker, D. J. (Eds.), {\it The Logico-Algebraic Approach to Quantum Mechanics}. Dordrecht: Springer (1979).

\bibitem{piron1976} Piron, C.: {\it Foundations of Quantum Physics}. Massachusetts: Benjamin (1976).

\bibitem{jauch1968} Jauch, J. M.: {\it Foundations of Quantum Mechanics}. Reading, Massachusetts: Addison-Wesley (1968).

\bibitem{varadarajan1968} Varadarajan, V.S.: {\it The Geometry of Quantum Theory}. New York: Springer-Verlag (1968).

\bibitem{piron1964} Piron, C.: Axiomatique quantique. {\it Helvetica Physica Acta}, {\bf 37}, 439-468 (1964).

\bibitem{mittelstaedt1963} Mittelstaedt, P.: {\it Philosophische Probleme der Modernen Physik}. Manheim: Bibliographisches Institut (1963). 

\bibitem{mackey1963} Mackey, G.: {\it Mathematical Foundations of Quantum Mechanics}. New York: Benjamin (1963).

\bibitem{mackey1957} Mackey, G.: Quantum mechanics and Hilbert Space. {\it American Mathematical Monthly}, {\bf 64}, 45-57 (1957).

\bibitem{birkhoffvonneumann1936} Birkhoff, G. and von Neumann, J.: The logic of quantum mechanics. {\it Annals of Mathematics}, {\bf 37}, 823-843 (1936).

\bibitem{tittelbrendelgisinherzogzbindengisin1998} Tittel, W., Brendel, J., Gisin, B., Herzog, T., Zbinden, H. and Gisin, N.: Experimental demonstration of quantum correlations over more than 10 km. {\it Physical Review A}, {\bf 57}, 3229-3232 (1998).

\bibitem{rosch1973} Rosch, E.: Natural categories, {\it Cognitive Psychology}, {\bf 4}, 328-350 (1973).

\bibitem{rosch1975} Rosch, E.: Cognitive representations of semantic categories. {\it Journal of Experimental Psychology: General}, {\bf 104},192-233 (1975).

\bibitem{chandler2002} Chandler, D.: {\it Semiotics: The Basics}. London: Routledge (2002).

\end{thebibliography}
\end{document}